\documentclass[prl,twocolumn,showpacs,floatfix,superscriptaddress]{revtex4-1}
\usepackage{amsmath}
\usepackage{graphicx}
\usepackage{dcolumn}
\usepackage{bm}
\usepackage{float}
\usepackage{mathrsfs}
\usepackage[cmyk]{xcolor}

\begin{document}

\title{Creating two-dimensional bright solitons in dipolar Bose-Einstein
condensates}
\author{Patrick K\"oberle}
\email{koeberle@itp1.uni-stuttgart.de}
\author{Damir Zajec}
\author{G\"unter Wunner}
\affiliation{Institut f\"ur Theoretische Physik 1, Universit\"at Stuttgart,
70550 Stuttgart, Germany}
\author{Boris A. Malomed}
\affiliation{Department of Physical Electronics, School of
Electrical Engineering, Faculty of Engineering, Tel Aviv University,
Tel Aviv 69978, Israel}
\date{\today}

\begin{abstract}
We propose a realistic experimental setup for creating quasi-two-dimensional
(2D) bright solitons in dipolar Bose-Einstein condensates (BECs), the
existence of which was proposed in Phys. Rev. Lett. \textbf{100}, 090406
(2008). A challenging feature of the expected solitons is their strong
inherent anisotropy, due to the necessary in-plane orientation of the local
moments in the dipolar gas. This may be the first chance of making
multidimensional matter-wave solitons, as well as solitons featuring the
anistropy due to their intrinsic dynamics. Our analysis is based on the
extended Gross-Pitaevskii equation, which includes three-body losses and
noise in the scattering length, induced by fluctuations of currents inducing
the necessary magnetic fields, which are factors crucial to the adequate
description of experimental conditions. By means of systematic 3D
simulations, we find a ramping scenario for the change of the scattering
length and trap frequencies which results in the creation of robust
solitons, that readily withstand the concomitant excitation of the
condensate.
\end{abstract}

\maketitle

Bose-Einstein condensation (BEC) of magnetic atoms, or molecules carrying
electric moments, has attracted a great deal of attention, starting from the
pioneering theoretical analyses of such condensates \cite%
{Goral2000,*Santos2000,*Goral2002}, which was followed by the creation of
BEC in the ultracold gas of $^{52}$Cr atoms, using magnetic \cite%
{Griesmaier2005,*Stuhler2005,*Griesmaier2007,*Koch2008} and all-optical \cite%
{Beaufils2008,*Pasquiou2011}\ techniques. Very recently, a condensate of $%
^{164}$Dy atoms, with a much larger magnetic moment, was created too \cite%
{Lu2010,*Lu2011}. The making of magnetic- and electric-dipolar BEC may also
be expected, respectively, in erbium \cite{McClelland2006} and in molecular
gases \cite{Deiglmayr2008,*Ospelkaus2010}. An updated review of this rapidly
progressing field was given in Ref. \cite{Lahaye2009}.

One of the promising possibilities, which has been, thus far, investigated
only theoretically, is the creation of bright solitons in dipolar
condensates. This can be easily predicted in various one-dimensional (1D)
settings, using, in particular, periodic potentials induced by optical
lattices as the stabilizing factor \cite%
{Zheng2007,*Gligoric2008,*Gligoric2009_1,*Gligoric2009_2,*Baizakov2009,*Cuevas2009,*Rojas2011,*Young2011}%
. In the limit of a very deep lattice, the BEC\ wave function becomes nearly
discrete, which makes it possible to predict stable 2D dipolar solitons \cite%
{Gligoric2010_1} and vortices \cite{Gligoric2010_2}, as well as 3D solitons
\cite{Muruganandam2011}, in the (quasi-)discrete form.

As concerns solitons, the most challenging issue is the creation of such
stable matter-wave modes in quasi-2D (pancake-shaped) condensates, as well
as their counterparts in the form of spatiotemporal solitons in optics. In
spite of intensive theoretical discussions \cite{Malomed2005,*Kartashov2011}, no experimental
results in this area have been reported thus far, the most essential
obstacle being the inherent instability of 2D solitons to the collapse,
driven by the self-focusing cubic nonlinearity (very recently, it was
proposed to create stable bright solitons by means of self-defocusing
nonlinearity, which is possible in a rather exotic situation with the
strength of the nonlinearity growing at $r\rightarrow \infty $ faster than $%
r^{D}$, where $D$ is the spatial dimension \cite{Borovkova2011_1,*Borovkova2011_2}). The dipolar
condensates suggest new challenges and possibilities for achieving this
fundamental purpose, by making use of the competition between local
interactions and long-range dipole-dipole interactions (DDIs). A quasi-2D
isotropic configuration implies that the local moments are polarized
perpendicular to the pancake's plane, in which case the DDI is repulsive. In
that case, the creation of (bright) 2D solitons may be possible if the sign
of the DDI is effectively reversed by means of a rapidly oscillating
magnetic field \cite{Pedri2005}; in the same setting, stable isotropic
solitons with embedded vorticity were predicted too \cite{Tikhonenkov2008_1}%
, and various 2D localized structures may be stabilized by trapping
potentials acting in the plane \cite{Lashkin2007}. The very fact of the
existence of multidimensional solitons in the dipolar BEC can be proven in a
rigorous mathematical form \cite{Antonelli2011}. It is also relevant to
mention that stable isotropic solitons are possible in optical media with
nonlocal (thermal or orientational) nonlinearities, as was demonstrated
experimentally \cite{Peccianti2002,*Conti2004,*Rotschild2005,*Rotschild2006}, and studied in detail theoretically \cite%
{Bang2002,*Krolikowski2004,*Briedis2005,*Buccoliero2007}.

The most challenging issue, which is unique to the multidimensional dipolar
media, is the possibility of the creation of a new species of  \emph{%
anisotropic} solitons, based on the in-plane polarization of local moments
(obviously, the anisotropy cannot manifest itself in 1D solitons). The
stationary form of such quasi-2D solitons, and some of their dynamical
properties, were investigated in Refs. \cite{Tikhonenkov2008_2,Eichler2011},
but the problem of the actual creation of the solitons under experimentally
relevant conditions has not been addressed before. The complex anisotropic
structure expected in the solitons makes this problem crucially important,
as the mode may be considered physically robust only if it is capable to
self-trap from experimentally feasible initial configurations. In this
work, we show, by means of systematic simulations of a realistic model,
that anisotropic solitons in dipolar BEC can be created, using available
experimental techniques. This result may have more general implications,
suggesting the realizability of various complex modes predicted in
multidimensional quantum gases, such as skyrmions \cite{Ruostekoski2001,*Battye2002,*Leanhardt2003,*Savage2003,*Kasamatsu2005}.

As mentioned above, there are currently two atomic species available for
creating magnetic dipolar BEC, \textit{viz}., $^{52}\mathrm{Cr}$ and $^{164}%
\mathrm{Dy}$. Since the latter has been condensed very lately \cite%
{Lu2010,*Lu2011}, our main results refer to chromium, but at the end of this
work we address dysprosium too. The making of the anisotropic solitons is
possible in a strongly dipolar regime, which may be defined by the
dimensionless ratio of the DDI and local interaction, $\epsilon _{\mathrm{DD}%
^{{}}}=m\mu _{0}\mu ^{2}/(12\pi a\hbar ^{2})$. Here, $m$ is the atom's mass,
$\mu _{0}$ the free-space permeability, $\mu $ the atom's magnetic moment,
and $a$ the $s$-wave scattering length. As shown in Ref.\ \cite%
{Tikhonenkov2008_2}, the necessary condition for the existence of solitons
is $\epsilon _{\mathrm{DD}^{{}}}>1$. In the case of $^{52}\mathrm{Cr}$, the
natural value of this ratio is $\epsilon _{\mathrm{DD}^{{}}}\approx 0.16$
\cite{Griesmaier2006}. Therefore, it is necessary to tune the scattering
length to enhance the relative strength of the DDI, which may be
accomplished by using the Feshbach resonance \cite{Werner2005,Lahaye2007}.
For $^{52}\mathrm{Cr}$, we thus find that the critical scattering length
below which the 2D solitons can exist is $15\,a_{\mathrm{B}^{{}}}$. However,
this condition is not sufficient, as it is only an upper bound, and the
soliton may actually be found at still smaller values of $a$\ \cite%
{Eichler2011}. On the other hand, if $a$ is too small, i.e.\ $\epsilon _{%
\mathrm{DD}^{{}}}$ exceeds a certain critical value, the condensate will
collapse. Therefore, solitons only exist within a certain range of values of
the scattering length. For typical trap frequencies and numbers of atoms in
the condensate, the relevant range for chromium is in a ballpark of $4\,a_{%
\mathrm{B}^{{}}}$ \cite{Eichler2011}.

To explore conditions for the creation of the anisotropic solitons, we
consider BEC of $N$ atoms with magnetic dipole moment $\mu $, which are
fully polarized by means of an external uniform magnetic field. At a
sufficiently low temperature, the dynamics of the condensate obey the
extended Gross-Pitaevskii equation (GPE):
\begin{equation}
\left[ \hat{h}_{0}(t)+\hat{h}_{\mathrm{int}}(t)+\hat{h}_{\mathrm{loss}}(t)%
\right] \Psi (\bm{r},t)=\mathrm{i}\hbar \partial _{t}\Psi (\bm{r},t),
\label{GPE}
\end{equation}%
with the single-particle Hamiltonian
\begin{equation}
\hat{h}_{0}(t)=-\frac{\hbar ^{2}}{2m}\Delta +\frac{m}{2}\left[ \omega
_{x}^{2}(t)x^{2}+\omega _{y}^{2}y^{2}+\omega _{z}^{2}(t)z^{2}\right] ,
\label{h_0}
\end{equation}%
the mean-field interaction Hamiltonian
\begin{align}
\hat{h}_{\mathrm{int}}(t)=& \frac{4\pi a(t)\hbar ^{2}}{m}\left\vert \Psi (%
\bm{r},t)\right\vert ^{2}  \notag  \label{h_int} \\
& +\frac{\mu _{0}\mu ^{2}}{4\pi }\int \mathrm{d}^{3}r^{\prime }\,\frac{%
1-3\cos ^{2}\vartheta ^{\prime }}{\left\vert \bm{r}-\bm{r}^{\prime
}\right\vert ^{3}}\left\vert \Psi (\bm{r}^{\prime },t)\right\vert ^{2},
\end{align}%
and the loss term
\begin{equation}
\hat{h}_{\mathrm{loss}}(t)=-\frac{\mathrm{i}\hbar L_{3}}{2}\left\vert \Psi (%
\bm{r},t)\right\vert ^{4}.  \label{h_loss}
\end{equation}%
Here, $\omega _{x,y,z}$ are the trap frequencies in the respective
directions, $L_{3}$ determines the rate of three-body losses, and $\Psi $ is
the mean-field wave function. The dipoles are polarized along the $z$-axis,
and it is assumed that the trap frequencies in $x$- and $z$-directions may
be time-dependent, along with the scattering length of the contact
interaction, $a$. We aim to consider the following scenario: Start with the
dipolar BEC at parameters for which the condensate is stable in the extended
state filling the entire 3D trap, and ramp down the scattering length and
frequencies $\omega _{x},$ $\omega _{z}$ simultaneously to values at which a
stable anisotropic soliton is expected. In the course of doing this, the
trap will be switched off completely in the $x$- and $z$-directions.
Obviously, this condensate-steering scenario will excite the condensate. We
will exploit this fact for proving the self-trapping nature of the DDIs,
since stable oscillations without the external trap are only possible if the
long-range interactions provide for the self-confinement. To make our model
realistic, we include losses via the imaginary-potential term (\ref{h_loss}%
), with an empirically known loss coefficient, $L_{3}=2\times 10^{-40}$ m$%
^{6}$/s \cite{Lahaye2008,Metz2009}, and random noise in the scattering
length, which stems from the current in the magnetic coils inducing the
Feshbach resonance. The latter is of special significance, because a typical
root-mean-square value of the noise, $\simeq 1-2~\,a_{\mathrm{B}^{{}}},$ is
not much smaller than the range of $a$ within which the soliton exists. The noise
that we used in the simulations is a uniform distribution of all frequencies
up to $500$ Hz. We do not expect higher frequencies to significantly affect
the dynamics of the soliton because the time scale of these dynamics is $%
\sim 1$ s, as shown below. We have analyzed the noise signal of the
Feshbach-inducing current in the experimental setup employed at the
laboratory in Stuttgart \cite{Henn2011}, observing close agreement with the
uniform distribution adopted in our scheme.

\begin{figure}[tbp]
\includegraphics[width=0.8\columnwidth]{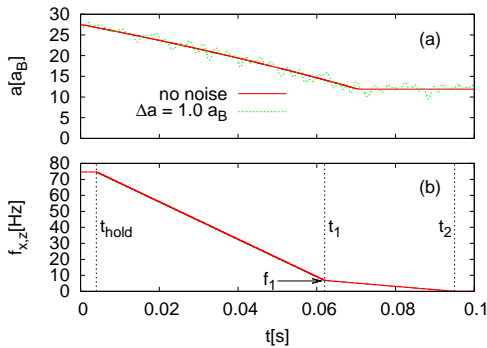}
\caption{(Color online) Ramping of (a) scattering length $a$ and (b) trap
frequencies. Only the first $100$ ms are shown. In the course of the
remaining time of the simulations, the scattering length is held at a
constant value, while the trap has already been switched off. In the
experiment, the scattering length cannot be held at an exact value, but
contains some noise, which is included into our scheme, as shown by the
green line in panel (a).}
\label{ramp}
\end{figure}

The ramping sequence is shown in Fig.\ \ref{ramp}. The initial condensate
contains $20000$ atoms, and is prepared in the trap with $(\omega
_{x},\omega _{y},\omega _{z})=2\pi \times (72,426,72)\,\mathrm{Hz}$ and $%
a=27\,a_{\mathrm{B}^{{}}}$. The scattering length is then ramped down to $%
11.5\,a_{\mathrm{B}^{{}}}$ within $70$ ms. The trap is first held constant
for $t_{\mathrm{hold}}=4$ ms and is then ramped to $f_{x,z}=f_{1}=6.5$ Hz
within $58$ ms. Finally, it is switched off within subsequent $33$ ms. We
emphasize the necessity for eliminating the trap. Even a very weak trapping
potential, with $f\sim $ a few Hz, affects the behavior of the soliton,
making it very difficult to distinguish the resulting dynamics from that of
an ordinary trapped condensate. For the simulations, the 3D spatial domain
was discretized into $512\times 128\times 512$ grid points, and marching in
time is performed using the Split-Operator method. The DDI integral is
evaluated with the aid of the Fourier convolution formula. Since this
numerical scheme is very demanding, the code was implemented using the
parallel computing architecture CUDA, and it was run on a modern graphics
card, enabling a very high degree of parallelization.

\begin{figure}[tbp]
\includegraphics[width=0.8\columnwidth]{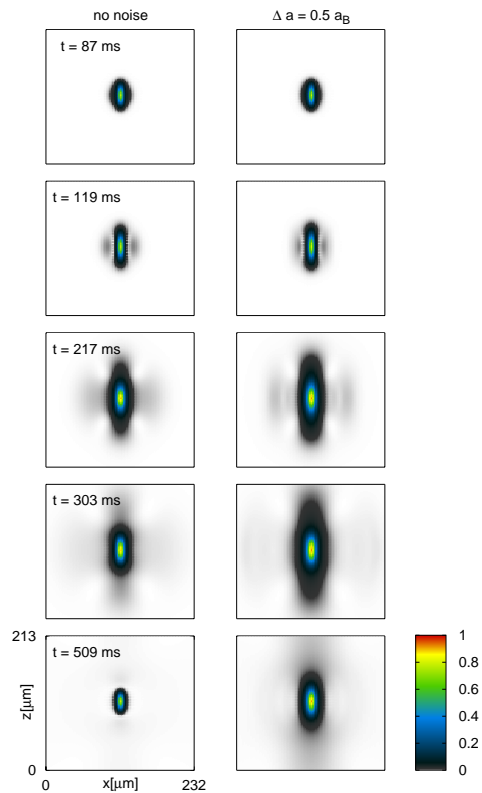}
\caption{(Color) The formation of a soliton. Left column: The scattering
length is ramped without the noise. Right column: The random noise with the
root-mean-square value of $\Delta a=0.5\,a_{\mathrm{B}^{{}}}$ was added to
the scattering length. Shown are absorption images of the condensate
density, with $\left\vert \Psi \right\vert ^{2}$ integrated along the $y$%
-axis and normalized to the maximum value. In both cases, the soliton is
being shaped in the first two frames, and is still expanding at $t=217$ ms.
Without the noise, the condensate self-focuses already at $t=303$ ms, while
this is slowed down when the noise is included. In both cases, the
refocusing occurs at the evolution stage when the external trap has already
been switched off completely, hence it is the DDIs which are responsible for
the self-trapping. This is a clear indication of the existence of a stable
excited soliton. It is important to note that the noise does not destroy the
soliton, which is crucial for the experiments. The field of the view is $%
232\,\protect\mu$m $\times 213\,\protect\mu$m.}
\label{sequence}
\end{figure}

Results of two such simulation runs are displayed in Fig.\ \ref{sequence}.
The left column shows absorption images of the condensate ($\left\vert \Psi
\right\vert ^{2}$ integrated over the $y$-direction and normalized to the
maximum value) in the absence of the noise in the scattering length, while,
in the right column, the noise with the root-mean-square value of $0.5\,a_{%
\mathrm{B}^{{}}}$ was added. In the first two frames, the soliton is being
shaped. It is observed that few atoms escape the core condensate in the $x$%
-direction, along which the DDI is repulsive. The condensate then grows
further in the $x$- and $z$-directions, but eventually inverts its dynamics
and refocuses. Without the noise, this occurs at $t\approx 303$ ms. Since
this is long after the trap has been switched off, the self-confining
potential needed for this outcome can only be provided by the DDIs. Thus, we
find a clear evidence for the self-trapping of the robust soliton.
Remarkably, the condensate refocuses as well when the noise is added to the
scattering length, as is shown in the right column of Fig.\ \ref{sequence},
i.e., the buildup of the soliton is robust against this experimentally
inevitable noise.

\begin{figure}[tbp]
\includegraphics[width=\columnwidth]{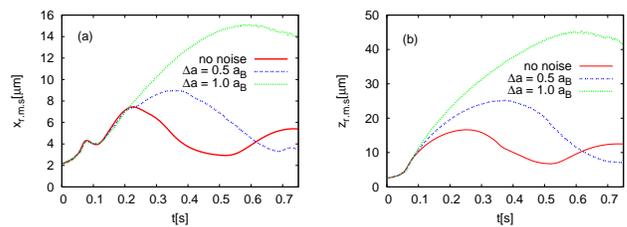}
\caption{(Color online) Root-mean-square extensions of the soliton in (a) $x$%
- and (b) $z$-directions for various values of the noise. Naturally, the
largest extension of the condensate increases when the noise is added to the
scattering length. At the end of the simulations performed without the
noise, the soliton has already started a second cycle of the oscillations.}
\label{extensions}
\end{figure}

To gain more insight into the dynamics, we monitored the root-mean-square
extensions of the condensate in the $x$- and $z$-directions during the
self-trapping process, as shown in Fig.\ \ref{extensions}. It is observed
that the expansion in these directions is roughly synchronous, which implies
that a breathing mode has been excited.

When the noise is added to the scattering length, the condensate reaches a
greater extension, and the oscillation frequency decreases. Above a critical
amplitude of the noise, for which we can give the lower bound of $1\,a_{%
\mathrm{B}^{{}}}$, we expect the soliton to decay. This behavior can be
understood in terms of a simple picture: One can replace the noise by an
effectively larger scattering length, which also lowers the binding energy
of the system and therefore slows down its dynamics. Consistent with Ref.
\cite{Eichler2011}, the increase of the scattering length, $a$, also leads
to a larger extension of the soliton. Above the critical value of $a$, the
soliton decays into a freely expanding condensate. From the experimental
point of view, the noise may even help to watch the soliton dynamics \textit{%
in situ} because of the comparably large observable change in the condensate
extensions.

The small dip in the $x$-extension at around $100$ ms is noteworthy, as it
is caused by the still present external trap with a small frequency, $f_{1}$%
, which decelerates the expansion. In line with this trend, increasing $%
f_{1} $ from $6.5$ Hz to $8$ Hz we have observed a more pronounced dip and a
smaller value of the maximum extension, together with an increased
oscillation frequency. For smaller holding times $t_{\mathrm{hold}}$,
meaning that the ramping sequence for the trap begins earlier, we find a
shallower dip, leading to slower dynamics and larger extensions of the
condensate.

\begin{figure}[tbp]
\includegraphics[width=0.7\columnwidth]{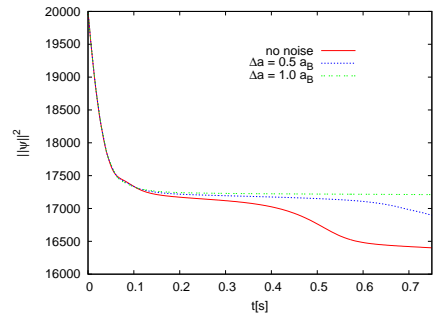}
\caption{(Color online) The number of atoms in the condensate in the course
of the formation of the soliton. Most atoms are lost during the first $50$
ms, when the density is still large and the soliton has not yet been formed
completely. When the soliton starts its first cycle of oscillations, the
loss rate rapidly drops, and at the end of the simulation, the condensate
still keeps more than $80\%$ of the initial atoms. For the noise amplitude
of $1\,a_{\mathrm{B}^{{}}}$, the losses are not visible at all after the
trap has been switched off.}
\label{part_number}
\end{figure}

The simulations show that the time scale governing the dynamics of the
soliton is on the order of $1$ s, which is relatively large for trapped
condensates. The main limitation for holding times is imposed by losses
caused by three-particle collisions, which is modeled in Eq. (\ref{GPE}) by
imaginary potential (\ref{h_loss}) proportional to the squared number
density. From Fig.\ \ref{extensions}, we observe that, in the case of the
ideal ramp, the soliton is larger in the $x$- and $z$-directions by a factor
of $4-5$ than the initial condensate (in the $y$-direction, the size is
approximately constant). This results in the density reduced by factor $%
\simeq 15-25$, hence the loss rate will rapidly drop when the trap is
switched off. Indeed, Fig.\ \ref{part_number}, where the evolution of the
number of atoms in the condensate during the simulation is plotted,
demonstrates that most atoms are lost within the first $50$ ms. Without the
noise in the scattering length, the atom number then stays nearly constant
for subsequent $300-400$ ms, and decreases at the end of the simulation,
when the soliton has reached its smallest extension, i.e.,\ it has returned
to the largest density. With the noise present, the mean extensions are even
larger, therefore the losses are reduced even more. For the $^{52}\mathrm{Cr}
$ condensate with the scattering length around $11\,a_{\mathrm{B}^{{}}}$, we
thus conclude that the atom losses will not cause decay of the soliton.

Finally, for parameters of the $^{164}\mathrm{Dy}$ condensate we find that
the threshold scattering length necessary for the existence of the
anisotropic solitons is $133\,a_{\mathrm{B}^{{}}}$. Since the actual
scattering length of Dy is estimated to be smaller than this value \cite%
{Lu2010,*Lu2011}, anisotropic solitons may be created in this condensate by
means of a simpler procedure, gradually switching off the trap only and
keeping the scattering length constant. Applying a simple Gaussian trial
wave function to evaluate the mean-field energy of a condensate which is
only trapped in $y$-direction, as was done in Refs.\ \cite%
{Tikhonenkov2008_2,Eichler2011}, we can estimate the binding energies of the
soliton in the $\left( x,z\right) $ plane for various scattering lengths.
This energy scale determines the frequency of oscillation of the excited
soliton, showing a crucial dependence on the scattering length. For the
confinement of $20000$ atoms in the $y$-direction, we choose a trap
frequency of $200$ Hz, and conclude that the relevant time scale varies from
$0.5$ s to $2$ s for the scattering length between $126\,a_{\mathrm{B}^{{}}}$
and $122\,a_{\mathrm{B}^{{}}}$. At $128\,a_{\mathrm{B}^{{}}}$, the time
scale already reaches a value $\simeq 10$ s.

In conclusion, we have performed systematic simulations of the realistic
model describing the evolution of the dipolar condensate en route to the
formation of the recently predicted new species of localized
multidimensional solitary modes, in the form of quasi-2D anisotropic
solitons. The model includes factors which are crucially important to the
adequate description of the experiment, such as the three-body losses and
Feshbach-induced noise in the scattering length. The scenario for the
formation of the solitons was elaborated, by gradually switching off the
trap and decreasing the scattering length. It is demonstrated that, in the
chromium and dysprosium BEC alike, the solitons may be safely formed,
surviving the inevitable excitation of the condensate. Further challenging
problems may be the making of soliton clusters, and the study of
interactions between anisotropic solitons. A more general implication of the
reported results is the possibility to elaborate similar dynamical scenarios
for the creation, under realistic conditions, of various complex
multidimensional patterns predicted by the analysis in quantum gases.

\acknowledgements

We thank Stefan M\"{u}ller, Emanuel Henn, Juliette Billy, Tilman Pfau and
Luis Santos for valuable discussions. The work of B.A.M. was supported, in
part, by the German-Israel Foundation through grant No. I-1024-2.7/2009.

\bibliographystyle{apsrev4-1}
\bibliography{literature2.bib}

\end{document}